\documentstyle[preprint,aps]{revtex}

\newcommand{\be}{\begin{equation}}
\newcommand{\ee}{\end{equation}}
\newcommand{\bea}{\begin{eqnarray}}
\newcommand{\eea}{\end{eqnarray}}

\begin{document}
\preprint
\widetext

%%\include{psfig}
%\twocolumn[\hsize\textwidth\columnwidth\hsize\csname @twocolumnfalse\endcsname 

%%%%%%%%%%%%%%%%%%%%%%%%%%%%%%
\title{Elementary Excitations in Magnetically Ordered Systems with Orbital
Degeneracy } 
\author{A. Joshi$^1$,  M. Ma $^1$, F. Mila$^2$, D. N. Shi$^{1,3}$, and F. C.
Zhang$^1$\\ } 
\address{
         $^1$Department of Physics,
University of Cincinnati, Cincinnati, OH 45221-0011\\
	$^2$ Laboratoire de Physique Quantique, Universite Paul Sabatier,
31062 Toulouse Cedex, FRANCE \\
	$^3$ College of Science, Nanjing Univ. of Aeronautics and
Astronautics, Nanjing, People's Republic of China 
}
\date{\today}
\maketitle
\widetext
\begin{abstract}
\noindent
The Holstein-Primakoff transformation is generalized to develop a quantum
flavor wave theory for spin systems with orbital degeneracy. Elementary
excitations of ordered ground states consist of spin, orbital , and
spin-orbital waves. Spin and spin-orbital waves couple to each other due to 
orbital anisotropy and Hund's rule, resulting in new modes observable by
inelastic neutron scattering. In the
$SU(4)$ limit, flavor waves are dispersionless along one or more
directions, and give rise to  quantum fluctuations of reduced dimensionality.
\end{abstract}

\noindent PACS number(s): 75.10.-b, 75.10.Jm, 75.30.Ds, 11.30.-j

%vskip2pc]

%\narrowtext

\newpage

The Hubbard model with a Hilbert space of only one atomic orbital per site
has been a popular model used to study strongly correlated electronic
systems. When the correlation $U$ is sufficiently large, the system 
undergoes a transition into a Mott insulator. In the
large $U$ limit,  the model reduces to the familiar $S=1/2$
Heisenberg antiferromagnetic Hamiltonian (HAH). 
In spite of the success and popularity of the single orbital Hubbard model,
it cannot, however, explain the magnetic behavior of many Mott insulators,
including ferromagnetism, magnetic ordering patterns, such as that in $V_2O_3$
~\cite{McWhan,Word,Bao}, which would require unrealistically large long-ranged
coupling if explained by HAH, and
paramagnetic behavior on lattices where the HAH is known to
exhibit long-range order (LRO) ~\cite{Takano}. On the other hand,
such behavior can be easily understood if
one allows for orbital degeneracy and enlarges the Hilbert space on each site
~\cite{Caste,Rice,Li}.  Recently, direct evidence for
orbital degrees of freedom and ordering in some magnetic systems have been
obtained by form factor analysis of x-ray diffraction ~\cite{Tokura}.  

In the case of spin systems with two-fold orbital degeneracy, the Hamiltonian 
is quadratic in operators $S_n^m$ , the generators of $SU(4)$ Lie algebra.  
Indeed the Hamiltonian may have 
$SU(4)$ symmetry~\cite{Li,Arovas}, rather than just the spin $SU(2)$
symmetry.  In many 2D lattices such as the square or triangular lattice with
nearest neighbor (n.n) coupling, the $SU(4)$ antiferromagnetic (AF)
Hamiltonian has no LRO even at $T=0$, as indicated by
mean field theory, variational calculations~\cite{Li}, and recent Monte Carlo
simulations~\cite{Tosatti}.  LRO may be stabilized in higher
dimensions. In real systems, the $SU(4)$ symmetry is at best approximate due
to orbital anisotropy and Hund's rule. With sufficient deviation from the
$SU(4)$ limit, LRO may be attained even in $2D$.

In this paper we investigate the elementary excitations of the orbital-spin
systems \textit{assuming }the ground state has broken symmetry. In effect,
we are studying for orbital-spin systems the equivalence of the familiar spin
waves, henceforth called flavor waves. For spin only systems, the
Holstein-Primakoff transformation (HPT) maps spin operators into boson
operators, and the linearized spin wave theory is equivalent to a
non-interacting boson problem. For orbital-spin systems, 
we show that the $SU(4)$ algebra can be exactly reproduced by a
generalized HPT involving three bosons ($N-1$
bosons for $SU(N))$. A 
quantum  flavor wave theory
is then developed. We find that in the $SU(4)$ limit, even though the
underlying lattice and ordering pattern is two-dimensional (2D), the flavor
wave excitations can be one-dimensional (dispersionless along one direction) or
localized (dispersionless in all directions). Quantum fluctuations of
these excitations give rise to disordering effect of reduced dimensionality
which provides further support for the lack of
LRO~\cite{Li,Tosatti}. 
The excitations in the
$SU(4)$ limit can be characterized as pure spin waves (spin rotation
only), pure orbital waves (orbital rotation only), and pure spin-orbital
waves (simultaneous spin and orbital rotation). The anisotropy and the Hund's
rule break the $SU(4)$ symmetry, and the spin and spin-orbital waves are in
general mixed.  We use a simplified model relevant to $V_2O_3$ to illustrate
the mixing, and predict new modes observable in 
neutron scattering
experiments.

With two orbital degrees of freedom, we  can define orbital operator 
$\vec \tau$
which acts on orbital states in the same way spin operator $\vec s$ on
spin states. The Hilbert space on each site consists of 4 basis states, which
we choose as $\mid s_{z},\tau _{z}\rangle .$ We label them as 
\begin{eqnarray}
&\mid &1\rangle =\mid \frac{1}{2},\frac{1}{2}\rangle ,\mid 2\rangle =\mid -%
\frac{1}{2},\frac{1}{2}\rangle ,  \nonumber \\
&\mid &3\rangle =\mid \frac{1}{2},-\frac{1}{2}\rangle ,\mid 4\rangle =\mid -%
\frac{1}{2},-\frac{1}{2}\rangle .  
\label{basis}
\end{eqnarray}
These basis states form a fundamental representation of $SU(4).$ The
conventional $SU(4)$ generators $S_{m}^{n}$ acts on the basis state $\mid
\l \rangle $ according to $S_{m}^{n}\mid l \rangle =\delta _{n,l }\mid
m\rangle $, and satisfies  $\sum_m S^m_m =1$ and $(S_m^n)^{\dagger} =S_n^m$.
The Lie algebra is given by 
\begin{equation} 
[S_{m}^{n},S_{k}^{l}]=\delta _{n,k}S_{m}^{l}-\delta _{m,l}S_{k}^{n}.
\label{SU4 commutator}
\end{equation}
$s_{\alpha },\tau _{\beta },$ and the product $s_{\alpha }\tau _{\beta }$
can all be expressed as linear combinations of the $S_{m}^{n}$~\cite{Li}.
For instance, $2s_z=S^1_1 - S^2_2 + S^3_3 - S^4_4$, $s^{\dagger} = S^2_1
+S^4_3$, and similarly for the orbital operators.  
Ignoring anisotropy and Hund's rule, the Hubbard model with double
orbital degeneracy in the large $U$ limit with 1 electron per site (1/4
filling) gives rise to the effective Hamiltonian  \begin{equation} 
H=\sum_{i,j}J_{ij}S_{m}^{n}(i)S_{n}^{m}(j), 
\label{SU4 Ham.} 
\end{equation}
with repeated indices $n,m$ summed. $H$ is clearly invariant under global \ $%
SU(4)$ transformation. We have argued~\cite{Li} that the ground state
of $H$ on many 2D lattices should be a flavor liquid. It is however useful
to study the possible excitations assuming the ground state has spontaneous
symmetry breaking. This is because i) while the ground state for the
fundamental representation is disordered, it may have LRO for higher
representations ii) such a study can provide information about the stability
of certain classical ground states against quantum fluctuations; and iii)
LRO can exist in $3D$ and/or away from the $SU(4)$ limit; and this analysis
will allow insight as to what aspects of those excitations are due to
proximity to $SU(4)$ symmetry.

Just as in the $SU(2)$ case, it is useful to generalize the $s=1/2$ problem
to general $s$,  we can also consider representations other than the
fundamental representation for $SU(4)$.  In particular, we generalize it to
representations denoted by Young tableaux with a single row but arbitrary
columns $M$. 
In the limit $M\rightarrow \infty $ , the non-commutativity between the
$S_{m}^{n}$ can be ignored, and the operators become \textrm{c}-numbers.
For general $M$, the Lie algebra (\ref{SU4 commutator}) can be exactly
reproduced by a generalized HPT using $3$-bosons $b^m_n$ at each
site, with $n \neq m$. The vacuum of these bosons is the state with
$S_{m}^{m}=M,$ where the choice of $m$ is formally arbitrary, but in practice
is taken to be the ordering ''direction'' of the classical ground state. The
generalized HPT is defined as (for $n,l \neq m$): 
\begin{eqnarray}
S_{m}^{m} &=&M-\sum_{n\neq m}b_{n}^{m\dagger }b_{n}^{m},  \nonumber \\
S_{n}^{m} &=&b_{n}^{m\dagger }\sqrt{M-\sum_{l\neq m}b_{l}^{m\dagger
}b_{l}^{m}}, \,  \nonumber \\
S_{n}^{l} &=&b_{n}^{m\dagger }b_{l}^{m}. \,  \label{Holstein}
\end{eqnarray}
Eq. (\ref{Holstein}) enables us to carry out the linear flavor wave
theory by expanding the Hamiltonian in power  of 1/M to  quadratic order in
$b$ and/or $b^{\dagger}$ and then set $M =1$  for the
present physical system.

First consider the ''ferromagnetic'' (FM) case,
$J_{ij}\leq 0$ in (\ref{SU4 Ham.}). The classical ground state, which is
also the exact ground state, is given by $S_{n}^{n}(i)=M \delta_{n,1}$.
Expansion of (\ref{SU4 Ham.}) leads to the flavor wave
Hamiltonian (omitting the superscript 1),
\[ H^{FM}_{fw}=\sum_{\langle ij \rangle,l}\left| J_{ij}\right| \left(
[ b_{l}^{\dagger }(i)b_{l}(i)-b_{l}^{\dagger }(i)b_{l}(j) ] +
i \leftrightarrow j \right) , \]
where $l=2,3,4$. This is comprised of  three independent boson
Hamiltonian corresponding to ''FM'' spin, orbital, and spin-orbital waves. 

For the AF $SU(4)$ system, the classical ground state may be
obtained by replacing the operators in (\ref{SU4 Ham.}) by their expectation
values with respect to states of the form $\left| \Psi \right\rangle
=\prod_{i}\left| \phi _{i}\right\rangle _{i},$ so that $\left\langle
S_{l}^{k}(r)\right\rangle =\left\langle \phi _{r}\left| S_{l}^{k}\right| \phi
_{r}\right\rangle ;$ and then minimizing the energy. Since
$\sum_{nm}\left\langle S_{n}^{m}(i)\right\rangle \left\langle
S_{m}^{n}(j)\right\rangle \geq 0,$ the classical minimum for each bond is
obtained by having the two sites connected satisfy $\left\langle \phi
_{i}\right| \left| \phi _{j}\right\rangle =0.$ This can be achieved by having
any two sites connected by a bond having different flavors. Assuming this can
be done for all bonds (unfrustrated), the classical ground state is then
identical to that of the AF $4$-state Pott's model. For
concreteness, let us consider for now the square lattice with n.n.  coupling
$J$ and next n.n. coupling $J^{\prime },$ with $J>2J^{\prime }$ . Out of the
degenerate manifold of classical ground states, we select for illustration the
one obtained by dividing the lattice into $4$ interpenetrating square
sublattices and having all sites on sublattice $\alpha $ be in the flavor
state $\left| \alpha \right\rangle $ (see Fig. 1a), i.e.$\left\langle
S_{n}^{m}(i)\right\rangle =\delta _{nm}\delta _{n\alpha _{i}},$ where $\alpha
_{i}$ denotes the sublattice the site $i$ is on. Expanding (\ref{SU4 Ham.})
to the leading order in $M$, we have $H= \sum_{mn} H_{mn}$, where $mn$ are
the pairs of the 4-states in (\ref{basis}) with $m \neq n$, and 
\begin{eqnarray}
H_{mn} = \sum_{i, j } J_{ij} \{ b^{m \dagger}_{n}(i) b^{m}_{n}(i) + 
b^{n \dagger}_{m}(j)b^{n}_{m}(j)  
+ [ b^{m \dagger}_n(i) b^{n \dagger}_m(j) + h.c.] \}, 
\label{spin wave Ham} 
\end{eqnarray} 
where $i$ and $j$ are summed over all the sites in sublattices $m$ and $n$
respectively.  Note that the different pairs of $H_{mn}$ are decoupled as a
consequence of the $SU(4)$ symmetry. A boson $b^m_n$ at sublattice $m$ only
coupled to its "mirror" boson $b^n_m$ at sublattice $n$.  This allows us to
solve each $H_{mn}$ separately, and $H_{mn}$ is  a simple Bogoliubov problem
identical to the usual spin wave theory in the HAH.  The
spin ($s$), orbital ($\tau$), and spin-orbital ($s\tau$) wave
dispersions are given by 
%the solutions of $H_{12}$ (or $H_{34}$), $H_{13}$ (or
%$H_{24}$), and $H_{14}$ (or $H_{23}$) respectively. They are  
$\omega_s (\vec k)=2J\sqrt{1-\cos^2 k_{y}}$,  $\omega_{\tau} (\vec
k)=2J\sqrt{1-\cos^2 k_{x}}$, and 
$\omega_{s\tau} (\vec k)=4J^{\prime}\sqrt{1-\cos^2k_{x} \, \cos^2k_{y}}$.  
Note that the
spin and the orbital excitations are 1D in spite of the underlying
translationally invariant 2D ordering pattern. The 1D density of states can be
 understood from the ordering pattern of the classical ground state in Fig.
1a.  There is no energy cost if we take any vertical or
horizontal line and interchange the flavors on the two sublattices (creating a
line defect). The disordering effect due to quantum fluctuations of these 1D
like excitations should act just like those in 1D quantum spin chains and
destroy the LRO. Of course, a linearlized theory can only be used as a guide
to the true situation. ~\cite{Mila,foot2}
Nevertheless the result here is consistent with the previous results that the
ground state of the AF $SU(4)$ model is a flavor liquid on many
2D lattices. In the limit $J^{\prime }\rightarrow 0,$ point defects ( e.g.,
flip flavor from $1$ to $4$ on a site) can be created with no cost in energy,
thus $\omega_{s\tau} \rightarrow 0$.    Note that for $J^{\prime }>0,$ the
classical ground state does not have finite entropy per site in the
thermodynamic limit, but the classical order is destroyed by inifnitesimal
quantum fluctuations (all $M<\infty )$ .

While the details of the above depend on the particular ordered state we
choose, the central results of no mixing between different waves
and existence of the excitations with reduced dimensionality are
consequences of the $SU(4)$ symmetry. We
now study the effects of deviation from this symmetry. 
We will illustrate the physics using an example which
is of particular interest, the 2D honeycomb lattice,  
corresponding to a single plane of V$_{2}$O$_{3},$ whose
magnetic behavior Castellani et. al. ~\cite{Caste} sought to explain by
invoking orbital degeneracy. The experimentally determined magnetic structure
of the ordered state is consistent with that of 4-sublattice ordering
discussed above with the lattice appropriately modified~\cite{Bao}, as shown in
Fig. 1b. The full Hamiltonian when Hund's rule and anisotropy are taken into
account is quite complicated. Since we are principally interested in how the
physics changes away from the $SU(4)$ symmetry in ways that are independent of
the precise model, we will ignore Hund's rule and consider a Hamiltonian
including  orbital anisotropy,
\begin{eqnarray}
H^{hc}=\frac{2}{U}\sum_{\left\langle ij\right\rangle }
S_{m}^{n}\left( i\right) S_{n^{\prime}}^{m^{\prime }}(j)
t_{nn^{\prime }}^{(ij)}t_{mm^{\prime }}^{(ij)},
\label{hc Ham}
\end{eqnarray}
where $\left\langle ij\right\rangle$ denotes n.n. pairs, 
$m,n,m^{\prime },n^{\prime }$ are summed from $1$ to
$4$, and $t^{(ij)}_{mn}$ is the hopping integral from state
$\mid m \rangle$ at site $i$ to state $\mid n \rangle$ at
site $j$~\cite{footnote}. Because of orbital anisotropy, 
$t_{mn}^{(ij)}$ is diagonal in spin space, but not  in orbital space. 
The hopping matrix for different orientational bonds  $\left\langle
i'j'\right\rangle$ and $\left\langle ij\right\rangle$ are related~\cite{Caste}
by a rotational transformation.  Note that while $t^{(ij)}_{mn}$ can be made
diagonal on any one bond, it cannot be simulataneously done for all bonds. We
choose the atomic orbitals with  the hopping matrix diagonal for the
''horizontal'' bonds with eigenvalues $t_{<}$ and $t_{>}$. The quantity  $\eta
=\frac{t_{>}-t_{<}}{t_{>}+t_{<}}$ is a measure of the strength of the
anisotropy. Expansion of the Hamiltonian in $M$ leads to the flavor wave
Hamiltonian,

\begin{eqnarray}
H^{hc}_{fw}=\frac{2}{U}\sum_{\left\langle ij \right\rangle} 
\sum_{\alpha \neq \alpha_i} [ b^{\alpha_i \dagger}_{\alpha}(i)
b^{\alpha_i}_{\alpha}(i) t^{(ij)}_{\alpha_i \alpha_j} 
t^{(ij)}_{\alpha \alpha_j} + \,
\sum_{\beta \neq \alpha_j} 
(b^{\alpha_i\dagger}_{\alpha}(i) b^{\alpha_j \dagger}_{\beta}(j)
t^{(ij)}_{\alpha_i \beta} t^{(ij)}_{\alpha_j \alpha} + h.c.) ] .
\label{flavor Ham} 
\end{eqnarray}

Consider first the $SU(4)$ limit, $t^{(ij)}_{mn}= t \delta_{m,n}$, namely $\eta
=0$.   $H^{hc}_{fw}$ is reduced to the form of (\ref{spin wave Ham}) with
$J=2t^2/U$, and  the spin,
orbital, and spin-orbital waves are decoupled. The different connectivity of
the honeycomb lattice and the ordering pattern in Fig. 1b results in some
modifications from that in the sqaure lattice in the dispersion of the
excitations. Flavors $1$ and $4$ (or $2$ and $3$) are connected as 
zig-zagging vertical chains, and spin-orbital waves are 1D like with
dispersion $\omega_{s \tau}(\vec k) = J\, \sin(\sqrt 3 |k_y|/2)$.
Flavors $1$ and $2$ (or $3$ and
$4)$ are unconnected, and spin waves are localized on a site and have zero
excitation energy, $\omega_s =0$.  Flavors $2$ and $4$ (or $1$ and $3$) are 
connected as two-site pairs which are
decoupled from other pairs. The orbital modes ($b_{1}^{3}$, $b_{2}^{4}$) are
thus also localized with  $\omega_{\tau} = 0$. 

With anisotropy, the situation changes significantly. Spin is still conserved,
but orbital no longer. Thus, spin modes ($b^1_2$, $b^3_4)$ and
spin-orbital modes ($b^1_4$, $b^3_2$) are now mixed, and therefore both
neutron  scattering active. However, for convenience, we continue to label
them as spin and spin-orbital modes. Orbital information can thus be obtained
indirectly from neutron scattering experiment. For $H^{hc},$ the dispersion of
the spin-orbital mode remains 1D like. The lack of coupling between
zig-zagging vertical chains can be understood simply as follows. Because the
hopping matrix is diagonal on the horizontal bonds,  the coupling between the
two sites connected by such a bond in (\ref{flavor Ham}) contains only
terms of the form  $S_{m}^{n}\left( i\right) S_{n}^{m}(j).$ Thus, this coupling
conserves $S_{n}^{n}(i)+S_{n}^{n}(j),$ and hence the chains are decoupled for
spin-orbital and also spin modes for the same reason as the $SU(4)$ case.
For spin wave mode, the dispersion in fact remains
dispersionless with zero excitation energy, indicating they remain localized,
albeit no longer on a single site. Pure orbital excitations remain  localized
on the two sites of each horizontal bond and decoupled from others. 

In Fig. 2a we show the energy dispersion as a function of $k_{y}$ for the
spin and spin-orbital modes, which are the modes that can be probed by
neutron scattering. Excitation energy is shown in units of $J=2t_>^2/U$. There
is no dispersion with $k_{x}$ since the chains are decoupled for these modes.
The four purely spin modes ($ b_2^1,b_1^2,b_4^3,b_3^4)$ in the $SU(4)$ limit  
remains dispersionless with anisotropy but with some spin-orbital
characteristics mixed in. The four spin-orbital modes($
b_{4}^{1},b_{3}^{2},b_{2}^{3},b_{1}^{4}),$ which for $SU(4)$ were all
degenerate, are split with anisotropy into two two-fold degenerate branches
with the upper branch no longer a Goldstone mode as $k_{y}\rightarrow 0.$ If
the effects of Hund's rule are included with effective strength $J_{H}$, chains
will be coupled, and all the modes will have 2D dispersion. Furthermore, any
accidental degeneracy and/or zero energy excitations will be removed. The
spectrum will consist of two degenerate Goldstone spin waves whose dispersion
is linear for small $k$ with a slope whose scale is determined by  $J_{H}$;
two other degenerate spin waves with a gap and dispersion also detemined by 
$J_{H}$; two degenerate spin-orbital waves with a gap and dispersion in $k_{x}$
of order $J_{H},$ but whose dispersion in $k_{y\text{ }}$is of order $J;$ two
other degenerate spin-orbital modes with gap of order $J\eta ,$ and $k_{x}$
and $k_{y}$ dispersion of order $J_{H}$ and $J$ respectively. The twofold
degeneracy in each case is required by the remaining $SU(2)$ spin symmetry of
the Hamitonian. These features are qualitatively in agreement with inelastic
neutron scattering data on $V_2O_3$ which shows that the effective
in-plane spin-spin coupling between parallel spins to be considerably
weaker than that between antiparallel
ones. Details of the calculation including Hund's rule and interplane coupling
of $V_{2}O_{3}$ will be discussed in a later publication.

Because of the mixing between spin and spin-orbital waves,
neutron scattering experiments will couple to all the modes mentioned above.
The intensity of inelastic neutron scattering cross section is proportional
to $Im \, \chi (\vec q,\omega )$,  where $\chi $ is the transeverse
susceptibility. Within the flavor wave approximation, the spin lowering 
operators $S^-(i) = S^1_2(i) +S^3_4(i)$, which is  $b^{\alpha_i
\dagger}_{\alpha_i+1}$ if $i \in$ sublattice $1$ or $3$, and 
$b^{\alpha_i}_{\alpha_i -1}$ if $i \in 2$ or $4$, and similarly for the
raising operator. The relative intensity can thus be
calculated straightforwardly using the Bogoliubov transformation. We show
an example of this for $T=0$ in Fig. 2b for $\eta =0.43$ and without Hund's
rule. For finite $T$, the intensity of each branch needs to be multiplied by
$(1+ 2/(e^{\beta \omega} -1))$.

In summary, the Holstein Primakoff transformation has been generalized to
develop a quantum flavor wave theory for spin systems with orbital
degeneracy.  In addition to spin and orbital bosonic excitations, proximity to
$SU(4)$ symmetry gives raie to a third mode of boson excitations observable by
neutron scattering. 

We thank W. Bao and I. Affelck for many interesting
discussions. The work is in part supported by DOE grant DE-FG03-98ER45687.

%\begin{figure} [htb]
%\epsfxsize=3.2in
%%\epsfysize=3.0in
%\epsffile{ma11.eps}  
\begin{figure}      
\caption{  Four-sublattice classical ground states considered in the text. 
a) Square lattice; \, b)  Honeycomb lattice.} 
\end{figure} 

%\begin{figure} [htb]
%\epsfxsize=5.0in
%\epsfysize=3.0in
%\epsffile{ma2.eps}  
\begin{figure}  
\caption{a) Spin (triangules) and spin-orbital (circles and squares) wave
spectra with orbital anisotropy $\eta = 0.43$ for the honeycomb lattice as a
function of $k$ for wavevector $(k_x, k)$.  Dashed curve shows  the (4-fold
degenerate) spin-orbital excitations in the $SU(4)$ limit. \, b) Calculated 
neutron scattering intensities (arbitray units) for the spin (triangles) and
spin-orbital (circles and squares) modes}. 
\end{figure}


\begin{references}
\bibitem{footnote} Eq. (6) may be derived from Hubbard model with
anisotropic hopping integrals~\cite{Caste}. Here we have dropped linear terms
in $S^m_n$, which may change the precise classical 4-sublattice ground state.
\end{references}
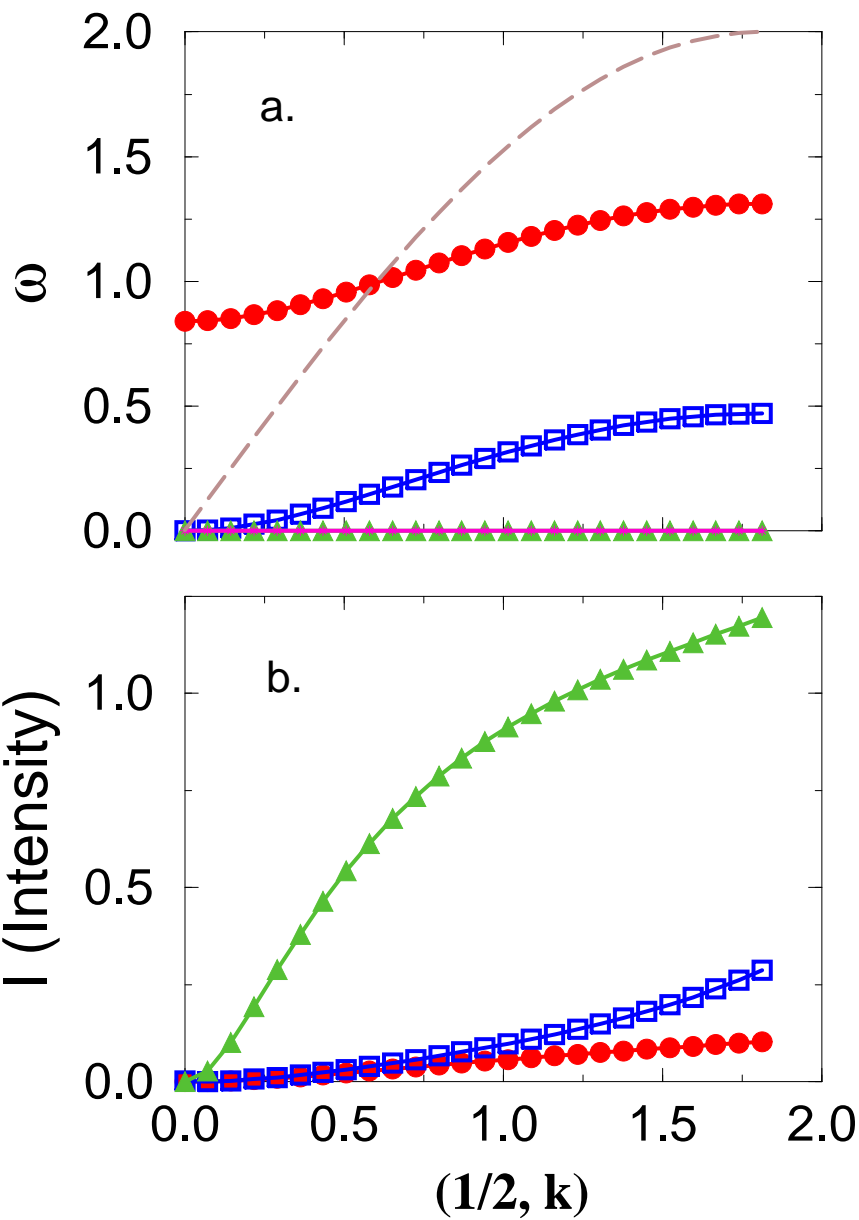
\end{document}